\documentclass[amsmath,amssymb,preprint]{revtex4-1}

\usepackage{graphicx}
\usepackage{bm}

\begin{document}

\title{Nonlinear Correlations in Multifractals: Visibility Graphs of Magnitude and Sign Series}

\author{Pouya Manshour}
\email{manshour@pgu.ac.ir}
\affiliation{Department of Physics, Faculty of Sciences, Persian Gulf University, 75169 Bushehr, Iran}

\begin{abstract}
Correlations in multifractal series have been investigated, extensively. Almost all approaches try to find scaling features of a given time series. However, the analysis of such scaling properties has some difficulties such as finding a proper scaling region. On the other hand, such correlation detection methods may be affected by the probability distribution function of the series. In this article, we apply the horizontal visibility graph algorithm to map stochastic time series into networks. By investigating the magnitude and sign of a multifractal time series, we show that one can detect linear as well as nonlinear correlations, even for situations that have been considered as uncorrelated noises by typical approaches like MFDFA. In this respect, we introduce a topological parameter that can well measure the strength of nonlinear correlations. This parameter is independent of the probability distribution function and calculated without the need to find any scaling region. Our findings may provide new insights about the multifractal analysis of time series in a variety of complex systems.
\end{abstract}

\maketitle


\section{Introduction}

Many biological, physical or social systems exhibit irregular behavior, which is a consequence of temporal or spacial interactions among their multitude components. The emergence of scale invariant properties is almost a common output of such complex systems, which are described by the theory of critical phenomena \cite{Wilson1971}. The fractal geometry \cite{Mandelbrot1982} has been widely used to study such scaling behaviors in various fields of researches such as physics, chemistry, biology, geology, neuroscience, engineering, finance, meteorology, and hydrology. Also, the fractal analysis helps us to better understand the underlying dynamics and to more precisely model such complex systems. 

If the fluctuations of all different magnitudes in a time series $x(t)$ scale with the same exponent, and some kind of homogeneity (linearity) exists in the behavior of the system over various scales from small to large ones, the system can be fully described by one parameter, called the Hurst exponent $H$ \cite{Hurst:1951aa}. This parameter characterizes the strength of linear correlation, and in stationary series indicates how fast the second order correlation function, $C(s)=\left\langle x(t+s)x(t)\right\rangle -\left\langle x(t)\right\rangle^2$, decays as a function of scale $s$. Any higher order ($>2$) correlation function can also be obtained by the second order one. This situation is called \textit{monofractality} \cite{Mandelbrot:2002aa}. However, this is not the whole story for many observed phenomena, i.e., often more than one exponent is needed to fully describe a complex system. The heterogeneity (nonlinearity) inherited in such systems results in the presence of different scaling behaviors for the fluctuations with different magnitudes, and higher order correlation functions are needed to describe system's features. This defines the concept of \textit{multifractality} \cite{Mandelbrot1999,Stanley1988}. Mono and multi fractality have been explored in a huge number of phenomena such as stock markets \cite{Matia2003,Oswieccimka2005}, turbulent flows \cite{Benzi1984,Meneveau1991}, earthquakes and seismic series \cite{Hirabayashi1992,Telesca2005,Manshour2009,Manshour2010}, human heartbeat dynamics \cite{Ivanov1999,Gieraltowski2012}, musics \cite{Su2006}, among others.

To analyze multifractal properties, various measures like generalized Hurst exponent, generalized dimension, scaling function, and multifractal spectrum have been introduced \cite{Kantelhardt2002,Lopes2009}. Among the numerous techniques that have been proposed to find such measures \cite{Lopes2009}, the multifractal detrended fluctuation analysis (MFDFA) has proved to be quite successful \cite{Kantelhardt2002}. Indeed, MFDFA is a generalization of the detrended fluctuation analysis (DFA) designed for monofractals \cite{Peng:1994aa}. In DFA method, after removing local trend in boxes of size $s$ by subtracting original series from a polynomial of a certain order, the root mean square (the second moment) of fluctuations, $F_2$, of the resulting series is obtained in all scales $s$. For power-law correlated series, we have $F_2(s)\sim s^{-\alpha}$, where $\alpha$ characterizes the strength of linear correlations in the series. In fact, for stationary ($\alpha<1$) and nonstationary ($\alpha>1$) linear correlated series $\alpha=H$  and $H+1$, respectively. In general, values of $\alpha<1/2$ represent anticorrelated series, and for $\alpha>1/2$ the series is positively correlated. At $\alpha=1/2$, the series is uncorrelated. It is worth to mention here that for a stationary and positively correlated linear series (i.e., $1/2<\alpha<1$), the second order correlation function scales as $C(s)\sim s^{-\gamma}$, with $\gamma=2-2\alpha$.

In multifractal series, MFDFA generalizes DFA method by analyzing the scaling of fluctuations for all moments of order $q$, that $q=2$ leads to DFA. For a power-law correlated series, the fluctuation function $F_q$ scales as $F_q(s)\sim s^{-H(q)}$, where $H(q)$ is the generalized Hurst exponent, from which we can find the scaling function $\tau(q)=qH(q)-1$, and the generalized dimension $D(q)=\tau(q)/(q-1)$. The multifractal spectrum $f(\alpha_q)$ indicates the distribution of scaling exponents $\alpha_q=d\tau/dq$, and can be obtained as $f(\alpha_q)=q\alpha_q-\tau(q)$. The width $\Delta\alpha_q$ of $f(\alpha_q)$ can be considered as a parameter for measuring the strength of multifractality. Note that for a monofractal linear series, $H(q)=\alpha$ is independent of $q$, and the multifractal spectrum becomes a delta function $f(\alpha_q)=\delta(\alpha_q-\alpha)$, thus $\Delta\alpha_q=0$. 

Another important approach for analyzing correlated processes that have been studied extensively is to decompose the series of increments ($x_i=X_{i+1}-X_i$) into magnitude ($x^{mag}_i=\vert x_i\vert$) and sign ($x^{sgn}_i=sign\left(x_i\right)$) series and then extract their scaling characteristics. For example, by using DFA and MFDFA methods, it has been shown \cite{ashkenazy:2001,Kalisky2005,Gomez2016} that the presence of correlation in the magnitude and sign series corresponds to the nonlinearity and linearity of the original series, respectively. This approach has been applied in various fields of study \cite{Ashkenazy2003,Zhu2012,Bartos2006,Li2014,Liu1999,Bernaola2017}.

In DFA and MFDFA techniques, one usually confronts with some challenges such as choosing an appropriate polynomial order for detrending procedure, finding a proper scaling region, and detecting correct correlations that can be affected by the probability distribution function (PDF) of the series. Recently, it has been shown that in some conditions, DFA and MFDFA are not able to extract correct scaling behaviors of a time series \cite{Gomez2016,Carpena2017,Bernaola2017}. The existence of crossovers in the scaling behavior at some particular scale $s_c$ along with the $q$ dependency of that $s_c$ are two examples of possible inaccuracy in the multifractal spectrum estimation. It is also indicated that in some situations MFDFA wrongly predicts linearity in spite of the nonlinearity of the studied time series. To correctly discover nonlinearity, they proposed to find the deviation of the second order correlation function of the series magnitude $C_{\vert x\vert}(s)=\left\langle \vert x(t+s)\vert \vert x(t)\vert\right\rangle -\left\langle \vert x(t)\vert\right\rangle^2$ from its expectation in linear Gaussian series \cite{Bernaola2017}. They argued that this method can be used for short series, and also can be applied even for series that do not show any scaling behavior. In a Gaussian multifractal model, they also showed that the nonlinearity implies the multifractallity, but the reverse is not true. 

In search of another possible ways for extracting correlations in time series, recently, Lacasa \textit{et. al.} introduced an algorithm, called visibility graph (VG) \cite{Lacasa2008,Luque2009,Lacasa2009}, that maps a time series into a graph based on the ability of the data points to see each other. In this approach, time series features are believed to be inherited in the resulting graph. For example, they showed that the monofractal exponent $H$ of a linear time series can be calculated from the degree distribution of the mapped graph. Thus, this algorithm may be considered as a novel method to analyze fractal and multifractal phenomena, along with other typical approaches like DFA and MFDFA. In spite of exploring various aspects of VG algorithm in different systems and situations \cite{Lacasa2010,Nunez2013,Lacasa2014,Lacasa2017,Manshour2015,Manshour:2015aa}, surprisingly, no general picture has emerged yet for multifractal series with nonlinear correlations. 

In this article, we apply the horizontal visibility graph algorithm to map fractal and multifratal time series into graphs. By investigating topological characteristics of the resulting graphs, we first show that this approach can well detect linear and nonlinear correlations, even for situations that DFA and MFDFA predict uncorrelatedness, due to some technical issues \cite{Carpena2017,Bernaola2017,Gomez2016}. On the other hand, we show that owing to the unique characterisitc of the horizontal visibility graph algorithm, one can calculate linear or nonlinear correlations, without the need to eliminate the impact of non-Gaussianity of the original series. Finally, we introduce a parameter that can well measure the strength of nonlinear correlation, where the multifractal spectrum width $\Delta\alpha_q$, that is a typical and widely used measure for such an analysis, is not able to discover such nonlinearities. Our results are in line with findings in recent studies \cite{Carpena2017,Bernaola2017}.

\section{Definitions: Simulated Series}
\label{MulSer}
We intend to investigate the impact of linear and nonlinear correlations as well as PDF of a series on the topological characteristics of the resulting visibility graph. In this respect, we generate series with adjustable these three features.

Fractional Brownian motion (fBm) was introduced to model a turbulent flow \cite{Kolmogorov:1940aa,Mandelbrot:1968aa}, and widely used in a variety of fields, including physics, statistics, hydrology, economy, biology, and many others \cite{Robinson:2003aa,Embrechts:2002aa,Mandelbrot:2002aa,Alvarez:2006aa,Varotsos:2006aa,Karagiannis:2004aa}. A fBm is a Gaussian monofractal process with stationary increments called \textit{fractional Gaussian noise} (fGn) and has long-memory which depends on the Hurst index, $H$ with $0<H<1$ \cite{Hurst:1951aa}. In fact, $H=1/2$ corresponds to the ordinary Brownian motion in which successive increments are statistically independent. For $H>1/2$, the increments are positively correlated, and for $H<1/2$, consecutive increments are more likely to have opposite signs or anti-correlated. One can generate such correlated series by using an algorithm called Fourier filtering method (FFM) \cite{Makse:1996aa}, as follows: multiply the Fourier transform of a generated white noise by a power-law of the form $f^{-\beta}$, and then Fourier transform the resulting series again in order to come back to the time domain. Finally, we have a correlated series with power spectrum of $S(f)\sim f^{-\beta}$. Note that $\beta=2\alpha-1$, where $\alpha$ is the DFA exponent.  

Both fGn and fBm series have Gaussian PDF. However, a wide range of natural and social phenomena exhibit a heavy-tailed PDF with infinite variance. To capture such heavy tails, various models have been proposed. Among them, the \textit{L{\'e}vy stable} distribution (LSD) has been considered extensively \cite{Tsallis1995,Mantegna1995}. The LSD is a family of all attractors of normalized sums of independent and identically distributed random variables. A symmetric LSD is characterized by the stability parameter $\lambda \in\left(0,2\right]$. For $0<\lambda\leq 1$, the distribution has an indefinite mean and variance value, and for $1<\lambda\leq 2$, it has a defined mean but infinite variance. The most well-known LSD functions are the Cauchy distribution with $\lambda=1$ and the Gaussian distribution function with $\lambda=2$. Thus, one can construct various uncorrelated series with non-Gaussian distributions, for different $\lambda<2$.

All time series defined above are linearly correlated (fBm and fGn) or completely uncorrelated (L{\'e}vy stable). To investigate the effect of nonlinear correlations, a \textit{multiplicative} multifractal series has been proposed by Kalisky \textit{et. al.} in \cite{Kalisky2005}. It can be generated by multiplying the magnitude and sign of two independent linear correlated time series with different DFA exponents of $\alpha_1$ and $\alpha_2$, as follows:
\begin{equation}
x_{mult}=\vert f_{\alpha_1}\vert sgn\left(f_{\alpha_2}\right)
\end{equation}
where $\vert ...\vert$ and $sgn\left(...\right)$ indicate the magnitude and sign operators. Also, $f_\alpha$ represents a correlated series generated by the Fourer filtering method described above with $\alpha=(\beta+1)/2$. One can control linear and nonlinear correlations in $x_{mult}$, with parameters $\alpha_2$ and $\alpha_1$, respectively (see \cite{Kalisky2005} for details).

\begin{figure}[t]
\begin{center}
\includegraphics[scale=0.6]{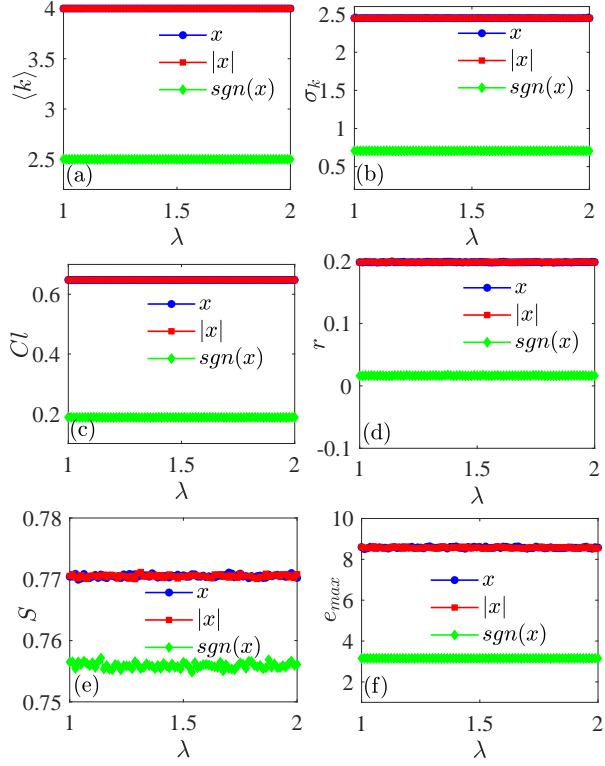}
\caption{The PDF dependency of (a) the average degree, $\left\langle k\right\rangle$, (b) standard deviation, $\sigma_k$, (c) clustering coefficient, $Cl$, (d) assortativity, $r$, (e) Spearman's coefficient, $S$, and (f) the maximum eigenvalue, $e_{max}$, of the adjacency matrix for original, $x$, magnitude, $\vert x\vert$, and sign, $sgn(x)$, series of a L{\'e}vy stable process with $\lambda\in [1,2]$. All features are completely independent of the non-Gaussianity of the series.}
\label{levy}
\end{center}
\end{figure}

\begin{figure}[t]
\begin{center}
\includegraphics[scale=0.6]{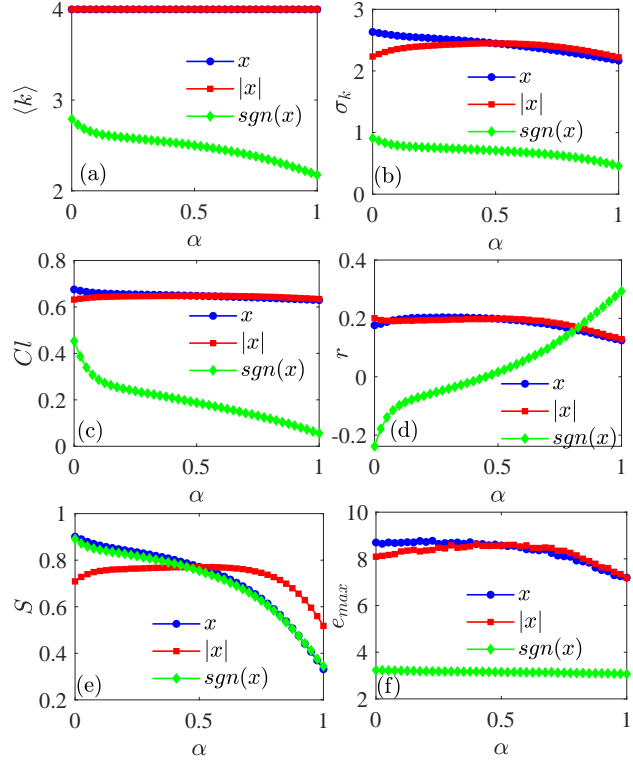}
\caption{The linear correlation dependency of (a) the average degree, $\left\langle k\right\rangle$, (b) standard deviation, $\sigma_k$, (c) clustering coefficient, $Cl$, (d) assortativity, $r$, (e) Spearman's coefficient, $S$, and (f) the maximum eigenvalue, $e_{max}$, of the adjacency matrix for original, $x$, magnitude, $\vert x\vert$, and sign, $sgn(x)$, series of a fractional Gaussian noise with $\alpha\in [0,1]$.}
\label{FFM}
\end{center}
\end{figure}

\section{Definitions: Visibility Graphs}
\label{HVG}
Two types of visibility algorithm have been introduced in \cite{Lacasa2008,Lacasa2009}. Here, we apply the horizontal visibility graphs (HVG) \cite{Luque2009}, which is a simpler algorithm with some advantages \cite{Manshour:2015aa} when compared with the (normal) visibility graph. Let $x_i$ be a series of $N$ data ($i=1,2,...,N$). By assigning each data point to a node in the graph, one can map a time series of size $N$ into a graph with $N$ nodes. Two nodes $i$ and $j$ are connected if one can draw a horizontal line in the time series joining $x_i$ and $x_j$ that does not intersect any intermediate data height, i.e., two arbitrary points $(t_i,x_i)$ and $(t_j,x_j)$ become two connected nodes, if any other data point $(t_q,x_q)$ placed between them satisfies $x_i,x_j>x_q$ for all $q$ such that $t_i<t_q<t_j$. It has been shown that HVG is always connected by definition and also is invariant under affine transformations, due to the mapping method. On the other hand, ordered and random series convert into regular and random exponential graphs, respectively. A unique characteristic of HVG is that this algorithm is independent of the PDF of the original series. It is worth to mention here that since PDF can affect the correlation estimation in time series, one usually need to replace original non-Gaussian PDF with a Gaussian one rank-wisely, to eliminate such distributional effects \cite{Schreiber2000,Manshour:2016aa}. But, it has been shown that this method performs well only where the data is linearly uncorrelated \cite{Ludescher2011}. We note that HVG has no such a limitation and thus has a special advantage in extracting correlation aspects of a time series, when compared with previous methods.

To analyze topological features of the mapped graphs, we first construct the corresponding adjacency matrix, $A$, so that $A_{ij}=1$ if nodes $i$ and $j$ are connected and $A_{ij}=0$, otherwise. The degree $k_i$ of an arbitrary node $i$ can be obtained via $k_i=\sum_jA_{ij}$. Afterward, the average degree $\left\langle k \right\rangle=\sum_kkp_k$ and standard deviation of the degree $\sigma^2_k=\sum_kk^2p_k-(\sum_kkp_k)^2$ can be calculated from the degree distribution $p_k$. To better investigate the effects of linear or nonlinear correlations on the resulting graphs, we will study some important topological features, defined in complex network theory. For example, the maximum eigenvalue, $e_{max}$, of the adjacency matrix is a key quantity in complex networks studies, and is proportional to the largest degree in networks with fat-tail degree distributions \cite{Chung2003}. The presence of any correlation between the degrees of pairs of connected nodes, called the assortativity $r$, is another important concept, studied extensively in complex networks \cite{boccaletti:2006}. The assortativity coefficient for mixing by node degree in an undirected network is 
\begin{equation}
r=\frac{\sum_{jl}jl(e_{jl}-q_jq_l)}{\sigma_q^2}
\end{equation}
where $e_{jl}$ is the fraction of edges that connect nodes of degrees $j$ and $l$, $q_k=(k+1)p_{k+1}/\left\langle k\right\rangle$, is the excess degree of a node defined as the number of edges leaving the node other than the one we arrived along, and $\sigma_q$ is the standard deviation of the distribution $q_k$. In general, we have $-1\leq r\leq 1$. Positive (negative) values of $r$ indicate a correlation between nodes of similar (different) degree. Also, if $r=1$, the network is said to be completely assortative, when $r=0$ the network is nonassortative, while at $r=-1$ the network is completely disassortative. 

The clustering coefficient, $Cl$, is also another important (three-point) correlation measure in complex networks, and can be considered as the density of connected triads of nodes in a network \cite{Watts:1998}. The local clustering of an arbitrary node $i$ is defined as $c_i=n_{e}/n_{p}$ where $n_{e}$ and $n_{p}$ represent the number of existing triads and the total number of possible triads, respectively. Thus, the average clustering coefficient of the network is $Cl=\sum c_i/N$. Finally, note that the time order of the original series, $x(t)$, is maintained in the resulting degree sequence, $k(t)=\{k_1,k_2,...,k_N\}$. This leads us to find the presence of any possible correlation between $x(t)$ and $k(t)$. In this respect, we calculate the Spearman correlation coefficient $S\in [-1,1]$, which measures the strength of a monotonic relationship \cite{spearman:1904}. 

In the next section, by using HVG algorithm we map our simulated series into networks, and then try to find topological characteristics mentioned above. We seek the effects of linear and nonlinear correlations as well as PDF of the original series on the resulting networks.

\begin{figure}[t]
\begin{center}
\includegraphics[scale=0.5]{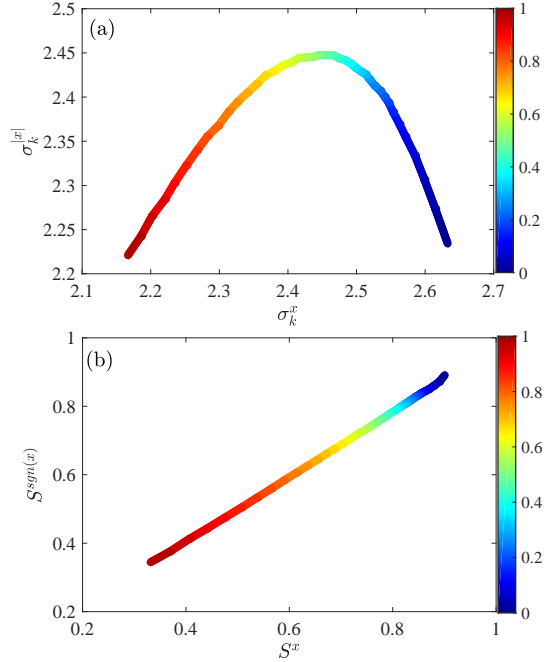}
\caption{The functionality of (a) $\sigma_k^{\vert x\vert}$ versus $\sigma_k^x$ and (b) $S^{sgn(x)}$ versus $S^x$. Colorbars show the corresponding values of the DFA exponent $\alpha$. The quadratic and linear behaviors observed in (a) and (b) are in complete agreement with recent studies \cite{Gomez2016,Bernaola2017}.}  
\label{sigma_S}
\end{center}
\end{figure}

\section{Results}

\begin{figure}[t]
\begin{center}
\includegraphics[scale=0.25]{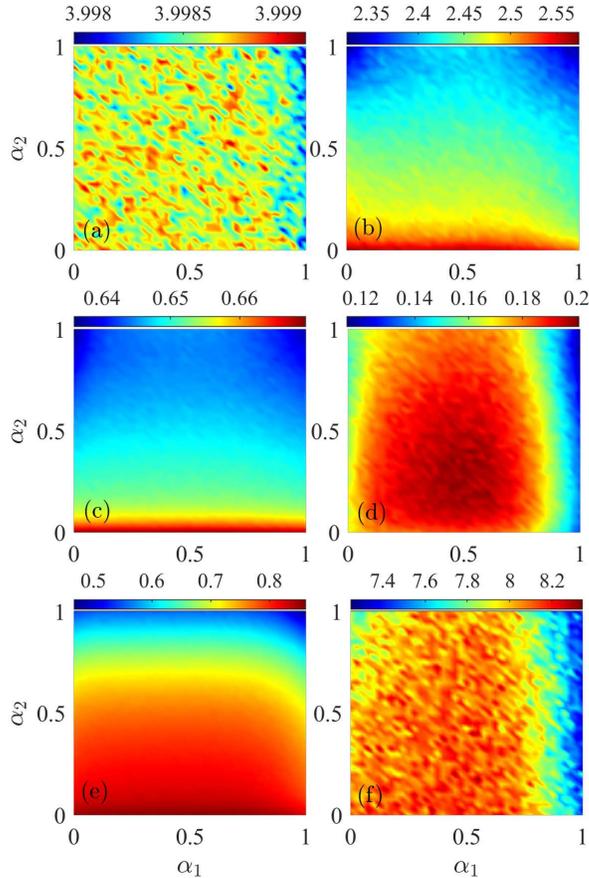}
\caption{The linear and nonlinear correlation dependency of (a) the average degree, $\left\langle k\right\rangle$, (b) standard deviation, $\sigma_k$, (c) clustering coefficient, $Cl$, (d) assortativity, $r$, (e) Spearman's coefficient, $S$, and (f) the maximum eigenvalue, $e_{max}$, of the adjacency matrix for original, $x$, magnitude, $\vert x\vert$, and sign, $sgn(x)$, series of a multiplicative series $x_{mult}$ with $\alpha_1,\alpha_2\in [0,1]$.}
\label{multipli}
\end{center}
\end{figure}

\begin{figure}[t]
\begin{center}
\includegraphics[scale=0.25]{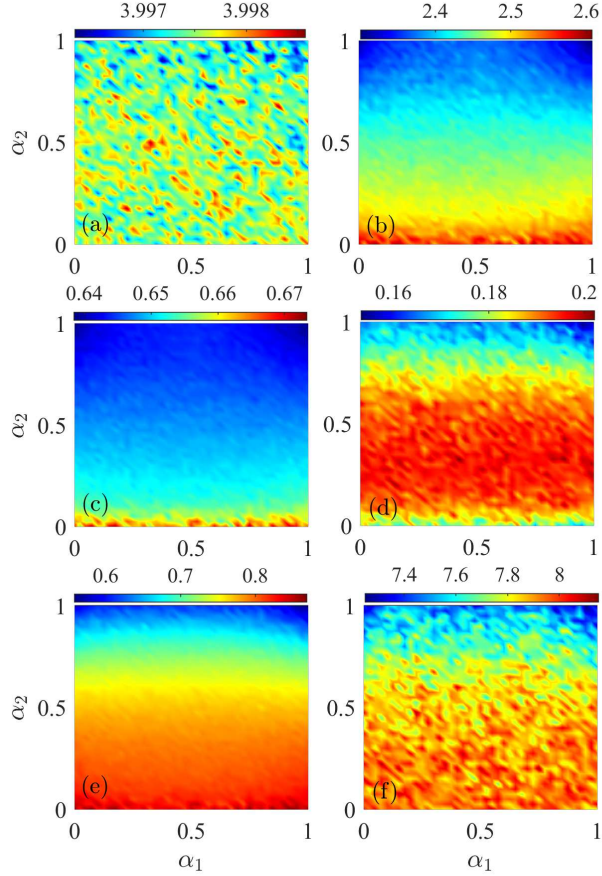}
\caption{The correlation dependency of (a) the average degree, $\left\langle k\right\rangle$, (b) standard deviation, $\sigma_k$, (c) clustering coefficient, $Cl$, (d) assortativity, $r$, (e) Spearman's coefficient, $S$, and (f) the maximum eigenvalue, $e_{max}$, of the adjacency matrix for original, $x$, magnitude, $\vert x\vert$, and sign, $sgn(x)$, series of a random-phased (RP) multiplicative process, $x^{RP}_{mult}$ with $\alpha_1,\alpha_2\in [0,1]$.}
\label{RP_comp}
\end{center}
\end{figure}

At first, to show that the horizontal visibility graph algorithm is independent of PDF of the original series, we plotted in Fig.~\ref{levy} six topological properties described in Sec.~\ref{HVG} corresponding to the original, $x$, magnitude, $\vert x\vert$, and sign, $sgn(x)$, of a L{\'e}vy stable process, with $1\leq \lambda\leq 2$. Fig.~\ref{levy}(a) to Fig.~\ref{levy}(f) represent the average degree $\left\langle k\right\rangle$, the standard deviation $\sigma_k$, clustering coefficient $Cl$, assortativity $r$, Spearman's coefficient $S$, and the maximum eigenvalue $e_{max}$ of the adjacency matrix, respectively. We note that by increasing $\lambda$ from $1$ to $2$, PDF transits from Cauchy (non-Gaussian) to Gaussian. As can be seen, non-Gaussianity cannot affect HVG and all plots are constants for various $\lambda$ values. We argue here that such a unique characteristic of HVG algorithm shows that without any additional action one can remove the PDF effects on the correlation properties of a given time series. Thus we can always apply this technique for situations where the exact estimation of correlations is the main goal. 

In order to investigate the impact of linear correlations on the HVG, we map fractal series with different linear correlations generated by $FFM$ methods for $0\leq\alpha\leq1$. The resulting topological features are plotted in Fig.~\ref{FFM}, similar to Fig.~\ref{levy}. Some parameters like $\sigma_k$, $Cl$, and $e_{max}$ can well discriminate positive and negative correlations in magnitude and original series. We note here that, DFA cannot detect any correlation in magnitude and sign series for $\alpha\leq 0.75$ and $\alpha\leq 1/2$, respectively \cite{Gomez2016}, and such regions are classified as uncorrelated white noises, wrongly. This result leads to another wrong conclusion that the presence of correlation in the magnitude is an indicator of nonlinearity in the original series. This actually is an spurious result of DFA and MFDFA \cite{Carpena2017,Bernaola2017}, and our findings confirm such studies, i.e., the original series here is a linear stochastic process, while its magnitude series is also correlated. Further, we find that average degree $\left\langle k\right\rangle$ of magnitude and original series are the same, and cannot distinguish series with different linear correlations. Also, the Spearman's coefficient for sign and original series are the same, now with a high power of discrimination between all values of $\alpha$. 

Based on the results depicted in Fig.~\ref{FFM}, it is interesting to find the functional form of $\sigma_k^{|x|}$ versus $\sigma_k^x$ as well as $S^{sgn(x)}$ versus $S^x$. Fig.~\ref{sigma_S} represents such relationships for various $\alpha$ values (colorbars). In Fig.~\ref{sigma_S}(a), we observe a symmetric behavior around $\alpha=1/2$, which shows that the magnitude series of a negatively correlated series is also positively correlated. Recently, it has been shown that for a linearly correlated Gaussian series, such a symmetric relationship is also observed in the behavior of the second order correlation function of the magnitude $C_{|x|}(s)$ versus original series $C(s)$ \cite{Bernaola2017}. We argue that the method used in \cite{Bernaola2017} is strongly dependent on the PDF of original series $x$, and thus before doing such an analysis one needs to replace rank-wisely the series values with a Gaussian ones, however in our approach no such a replacement is needed. Fig.~\ref{sigma_S}(b) also shows $S^{sgn(x)}$ versus $S^x$, and indicates a nearly complete linear relationship, showing a one by one correspondence between linear correlations in original and sign series. This result is in complete agreement with recent studies \cite{Gomez2016}.

To investigate the effects of nonlinearity on visibility graphs, we map a nonlinear multiplicative series, $x_{mult}$, described in section~\ref{MulSer} into HVGs for various values of $\alpha_1,\alpha_2\in [0,1]$. We calculate six topological features (similar to Figs.~\ref{levy} and \ref{FFM}) for various $\alpha_1$ and $\alpha_2$, in Fig.~\ref{multipli}. Colorbars in Fig.~\ref{multipli}(a) to Fig.~\ref{multipli}(f) show the values of $\left\langle k\right\rangle$, $\sigma_k$, $Cl$, $r$, $S$, and $e_{max}$, respectively. As can be seen, $\left\langle k\right\rangle$ and $e_{max}$ do not depend on linear correlations, and have a weak dependency on nonlinear correlations. On the other hand, $S$ is strongly dependent on linearity with a very weak dependency on nonlinear correlations for large $\alpha_1$. $r$ is strongly dependent on nonlinear correlations, but weakly depends on linearity, and $\sigma_k$ is dependent on linear as well as nonlinear correlations. Our results show that linear and nonlinear correltions are inherited in the HVG structure. To better understand this issue, we note that in linear time series, Fourier phases are completely random. Thus, a straightforward method to eliminate any nonlinear correlation is that after Fourier transforming a series $x$, we can shuffle its corresponding Fourier phases and then transform it back to the time domain to generate a phase randomized series $x^{RP}$ \cite{Schreiber2000}. In this respect, we eliminate nonlinear correlations in the multiplicative series $x_{mult}$ by shuffling its Fourier phases and then calculate topological characteristics of the mapped series $x_{mult}^{RP}$ in Fig.~\ref{RP_comp}. As expected, all features have lost their dependencies on nonlinear correlations (see Fig.~\ref{multipli} for comparison). This result confirms that the nonlinearity of $x_{mult}$ is only determined by $\alpha_1$.

\begin{figure}[t]
\begin{center}
\includegraphics[scale=0.2]{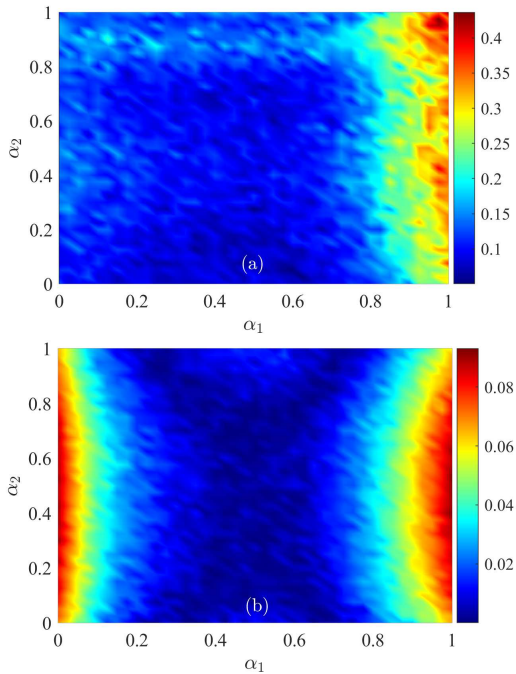}
\caption{(a) The multifractal width $\Delta\alpha_q$ calculated by MFDFA method for multiplicative series $x_{mult}$ with various $\alpha_1,\alpha_2\in [0,1]$ and (b) the nonlinearity measure $\Delta\sigma$ introduced in \ref{d_sig}, for the same series similar to (a).}
\label{new_measure}
\end{center}
\end{figure}

Now we seek to find a proper parameter that can well measure the strength of nonlinear correlations in a given series. Among these features, we choose $\sigma_k$ since it can discover positive as well as negative correlations in both original and magnitude series (see Fig.~\ref{FFM}). At first, we apply MFDFA to calculate the width $\Delta\alpha_q$ of the multifractal spectrum $f(\alpha_q)$. As we discussed in the introduction, this quantity is a typical parameter used to measure the strength of nonlinearity in a series. In Fig.~\ref{new_measure}(a) we plotted $\Delta\alpha_q$ for a multiplicative series with various $\alpha_1$ and $\alpha_2$. We observe that the series is nonlinear only for values of $\alpha_1>0.75$ and this nonlinearity increases with increasing $\alpha_1$. However, this is a sporious result of MFDFA discussed in \cite{Gomez2016,Bernaola2017}, extensively.

To solve this, we define a new topological parameter for measuring the strength of nonlinearity as follow:
\begin{equation}
\Delta\sigma=\frac{\vert \sigma_k^{|x|}-\sigma_k^{|x^{RP}|} \vert}{\sigma_k^{|x^{RP}|}}
\label{d_sig}
\end{equation}
where $x^{RP}$ shows the series $x$ that its nonlinearity has been destroyed with phase randomizing method. We plot $\Delta\sigma$ for multiplicative series $x=x_{mult}$ with various $\alpha_1$ and $\alpha_2$ in Fig.~\ref{new_measure}(b). Interestingly, we observe that this measure can well discover nonlinear features of the series where MFDFA is unable to detect any correlations, i.e., for all values of $\alpha_1<0.75$. On the other hand, a nearly symmetric behavior is observed around $\alpha_1=1/2$, which means that bellow and above this point, the strength of nonlinear correlation of the series is the same. Our results here are in line with results found in \cite{Gomez2016,Bernaola2017}, recently.

\section{conclusion}
\label{Conc}

Measurement of nonlinear correlations in stochastic time series is generally a difficult task. Many methods have been proposed to extract such information, among which DFA and MFDFA are of practical importance. All such methods are engaged with some challenges like finding a proper scaling region and eliminating the impacts of PDF for estimating the correlations, correctly. On the other hand, DFA and MFDFA may wrongly predict that a given correlated series is uncorrelated, which is due to some technical issues. Therefore, the search for possible new methods that can better analyze correlated time series is necessary. In this article, by using a recently proposed algorithm called horizontal visibility graph (HVG) that map a series into a graph, we investigate linear and nonlinear correlations in fractal and multifractal stochastic time series. Since HVG does not depend on the PDF of the original series, the resulting graphs contain only correlation information of the original series. We demonstrated that this unique feature can play the role of typical methods for eliminating the impacts of PDF, in which one usually replace the series values with a Guassian ones rank-wisely. We should note that such a replacement method only works for series with zero linear correlation. However, there is no such a restriction on the HVG algorithm, and thus it indicates the supremacy of our approach when the main aim is to estimate correlations, correctly. Further, we found that linear and nonlinear correlations are well inherited in the topological features of the resulting graphs. We note that this occurs even for time series in which DFA and MFDFA cannot detect any correlations. We also represented that the presence of correlations in series magnitude is not the indication of nonlinearity in the original series. At the end, we also introduced a topological parameter that can well measure the strength of nonlinearity. All such results obtain without the need to find scaling regions and demonstrate the unique power of correlation analysis via HVG algorithm. Consequently, our approach may be considered as a novel and precise method to estimate nonlinear correlations in various complex systems.

\begin{acknowledgments}
The support from Persian Gulf University Research Council is kindly acknowledged.
\end{acknowledgments}

\end{document}